\documentclass[conference]{IEEEtran}
\IEEEoverridecommandlockouts

\usepackage{tabularx}
\usepackage{array}
\usepackage{cite}
\usepackage{orcidlink}
\usepackage{amsmath,amssymb,amsfonts}
\usepackage{algorithmic}
\usepackage{graphicx}
\usepackage{textcomp}
\usepackage{subcaption}
\usepackage{xcolor}
\usepackage{amsmath}
\usepackage{algorithm}
\usepackage{algorithmic}
\usepackage{acronym}
\usepackage[inline]{enumitem}

\newcommand{\sectionname}{Sect.}
\renewcommand{\tablename}{Table}
\newcommand{\meter}{\,\mathrm{m}}
\newcommand{\mW}{\,\mathrm{mW}}
\newcommand{\dB}{\,\mathrm{dB}}
\newcommand{\dBm}{\,\mathrm{dBm}}
\newcommand{\MHz}{\,\mathrm{MHz}}

\newcommand{\bpsHz}{\,\mathrm{b/s/Hz}}

\newcommand{\Htran}{\mathsf{H}}
\newcommand{\Ttran}{\mathsf{T}}

\usepackage{tikz}
\usetikzlibrary{shapes, arrows.meta, positioning}

\tikzstyle{block} = [rectangle, draw, fill=blue!15,
    text centered, minimum height=1cm, minimum width=2.8cm]
\tikzstyle{line} = [draw, -{Latex[length=2mm]}]

\acrodef{2D}{bidimensional}
\acrodef{AWGN}{additive white Gaussian noise}
\acrodef{AdamW}{adaptive moment estimation with weight decay}
\acrodef{B5G}{beyond-5G}
\acrodef{CSA}{central satellite}
\acrodef{CFmMIMO}{cell-free massive \acs{MIMO}}
\acrodef{mMIMO}{massive \acs{MIMO}}
\acrodef{DPD} {disk Poisson Process} 
\acrodef{DRA}{direct radiating antenna}
\acrodef{DC}{direct current}
\acrodef{DoD}{depth of discharge}
\acrodef{eMBB}{enhanced mobile broadband}
\acrodef{E-GPW}{extended gridded population of world database}
\acrodef{FDM}{frequency division multiplexing}
\acrodef{FIR}{finite impulse response}
\acrodef{TFoA}{thinned \acs{FoA}}
\acrodef{FoA}{formation of arrays}
\acrodef{GEO}{geostationary Earth orbit}
\acrodef{GSO}{geosynchronous Earth orbit}
\acrodef{H-RRM}{heuristic radio resource management}
\acrodef{KPI}{key performance indicator}
\acrodef{IF}{intermediate frequency}
\acrodef{LEO}{low Earth orbit}
\acrodef{LoS}{line-of-sight}
\acrodef{MAI}{multiple access interference}
\acrodef{MB}{multi beam}
\acrodef{MEO}{medium Earth orbit}
\acrodef{MLP}{multi-layer perceptron}
\acrodef{mMTC}{massive machine-type communications}
\acrodef{eMTC}{enhanced machine-type communications}
\acrodef{PDD}{Poisson disk distribution}
\acrodef{EIRP}{effective isotropic radiated power}
\acrodef{NTN}{non terrestrial network}
\acrodef{URLLC}{ultra-reliable low-latency communications}
\acrodef{ECEF}{Earth-centered Earth-fixed}
\acrodef{GPS}{global positioning system}
\acrodef{HTFS}{high throughput fractionated satellite}
\acrodef{UHF}{ultra-high frequency}
\acrodef{FF}{formation flying}
\acrodef{MIMO}{multiple input multiple output}
\acrodef{M-MIMO}{massive multiple input multiple output}
\acrodef{GNC}{guidance navigation and control}
\acrodef{GNSS}{global navigation satellite system}
\acrodef{OFDM}{orthogonal frequency division multiplexing}
\acrodef{PHY}{physical-layer}
\acrodef{SA}{sub-array}
\acrodef{SADM}{solar array drive mechanism}
\acrodef{SG}{solar generator}
\acrodef{SNR}{signal-to-noise ratio}
\acrodef{SIR}{signal-to-interference ratio}
\acrodef{SINR}{signal-to-interference-plus-noise ratio}
\acrodef{SSPA}{solid-state power amplifier}
\acrodef{RF}{radio frequency}
\acrodef{R-GEO}{regional \acs{GEO}}
\acrodef{UT}{user terminal}
\acrodef{UE}{user equipment}
\acrodef{QVD}{quantized virtual distancing}
\acrodef{TDM}{time division multiplexing}
\acrodef{R-GEO}{regional GEO}
\acrodef{RRM}{radio resource management}
\acrodef{SSPA}{solid-state power amplifier}
\acrodef{BLER}{block error rate}
\acrodef{DL}{downlink}
\acrodef{UL}{uplink}
\acrodef{PSD}{power spectral density}
\acrodef{UDSM}{ultra deep sub-micron}
\acrodef{PEPS}{power energy platform simulation}
\acrodef{SE}{spectral efficiency}
\acrodef{wrt}{with respect to}
\acrodef{OBP}{on-board digital processor}
\acrodef{UDSM}{ultra-deep-sub-micron}

\acrodef{3GPP}{third generation partnership project}
\acrodef{AWGN}{additive white Gaussian noise}
\acrodef{B5G}{beyond-5G}
\acrodef{CS}{central satellite}
\acrodef{DRA}{direct radiating antenna}
\acrodef{DC}{direct current}
\acrodef{DoD}{depth of discharge}
\acrodef{eMBB}{enhanced mobile broadband}
\acrodef{FDM}{frequency division multiplexing}
\acrodef{FIR}{finite impulse response}
\acrodef{TFoA}{thinned \acs{FoA}}
\acrodef{FoA}{formation of arrays}
\acrodef{GEO}{geostationary Earth orbit}
\acrodef{GSO}{geosynchronous Earth orbit}
\acrodef{KPI}{key performance indicator}
\acrodef{IF}{intermediate frequency}
\acrodef{LEO}{low Earth orbit}
\acrodef{LoS}{line-of-sight}
\acrodef{MAI}{multiple access interference}
\acrodef{MEO}{medium Earth orbit}
\acrodef{mMTC}{massive machine-type communications}
\acrodef{eMTC}{enhanced machine-type communications}
\acrodef{EIRP}{effective isotropic radiated power}
\acrodef{NTN}{non terrestrial network}
\acrodef{URLLC}{ultra-reliable low-latency communications}
\acrodef{ECEF}{Earth-centered Earth-fixed}
\acrodef{GPS}{global positioning system}
\acrodef{HTFS}{high throughput fractionated satellite}
\acrodef{HTS}{high throughput satellite}
\acrodef{UHF}{ultra-high frequency}
\acrodef{FF}{formation flying}
\acrodef{MIMO}{multiple input multiple output}
\acrodef{GNC}{guidance navigation and control}
\acrodef{NTN}{non-terrestrial network}
\acrodef{GNSS}{global navigation satellite system}
\acrodef{OFDM}{orthogonal frequency division multiplexing}
\acrodef{PHY}{physical-layer}
\acrodef{SA}{satellite array}
\acrodef{SADM}{solar array drive mechanism}
\acrodef{SG}{solar generator}
\acrodef{SNR}{signal-to-noise ratio}
\acrodef{SIR}{signal-to-interference ratio}
\acrodef{SINR}{signal-to-interference-plus-noise ratio}
\acrodef{SSPA}{solid-state power amplifier}
\acrodef{RF}{radio frequency}
\acrodef{R-GEO}{regional \acs{GEO}}
\acrodef{UT}{user terminal}
\acrodef{UE}{user equipment}
\acrodef{TDM}{time division multiplexing}
\acrodef{RRM}{radio resource management}
\acrodef{SSPA}{solid-state power amplifier}
\acrodef{BLER}{block error rate}
\acrodef{DL}{downlink}
\acrodef{UL}{uplink}
\acrodef{RV}{random variable}
\acrodef{PSD}{power spectral density}
\acrodef{UDSM}{ultra deep sub-micron}
\acrodef{PEPS}{power energy platform simulation}
\acrodef{SE}{spectral efficiency}
\acrodef{wrt}{with respect to}
\acrodef{OBP}{on-board digital processor}
\acrodef{UDSM}{ultra-deep-sub-micron}
\acrodef{UC}{user-centric}
\acrodef{CSI}{channel state information}
\acrodef{PAC}{per-antenna constraint}
\acrodef{FPAC}{fair \acs{PAC}}
\acrodef{MPC}{maximum power constraint}
\acrodef{AWGN}{additive white Gaussian noise}
\acrodef{TX}{transmit}
\acrodef{MMSE}{minimum mean square error}
\acrodef{MF}{matched filter}
\acrodef{ZF}{zero forcing}
\acrodef{ELSA}{enhanced logarithmic spiral array}
\acrodef{UPA}{uniform planar array}
\acrodef{NB}{narrowband}
\acrodef{WB}{wideband}
\acrodef{BFN}{beamforming network}
\acrodef{MD-MIQP}{minimum distance mixed integer quadratic problem}
\acrodef{PDF}{probability density function}
\acrodef{NPR}{noise-to-power ratio}
\acrodef{HPA}{high power amplifier}
\acrodef{CF}{cell-free}
\acrodef{3GPP}{third generation partnership project}
\acrodef{UC-MIMO}{user centric MIMO}
\acrodef{VHTS}{very high throughput satellites}
\acrodef{PM-MIMO}{pragmatic M-MIMO}
\acrodef{NR}{new radio}
\acrodef{LMS}{land mobile satellite}
\acrodef{IM}{intermodulation}
\acrodef{OBO}{output back-off}
\acrodef{FDD}{frequency division duplexing}
\acrodef{TDD}{time division duplexing}
\acrodef{BLER}{block error rate}
\acrodef{rv}{random variable}
\acrodef{COB}{center of beam}
\acrodef{ISL}{inter-satellite link}
\acrodef{AP}{access point}
\acrodef{SVD}{single value decomposition}
\acrodef{CW}{continuous wave}
\acrodef{AOCS}{attitude and orbit control system}
\acrodef{LSTM}{long short-term memory}
\acrodef{CPU}{central processing unit}
\acrodef{GPU}{graphics processing unit}
\acrodef{TPU}{tensor processing unit}
\acrodef{MMF}{max-min fairness}
\acrodef{MHA}{multi-head attention}
\acrodef{FFN}{feed-forward network}
\acrodef{MSE}{mean square error}
\acrodef{CDF}{cumulative distribution function}
\acrodef{EPA}{equal power allocation}
\acrodef{FPA}{fractional power allocation}
\acrodef{TNN}{transformer neural network}
\acrodef{RT}{real-time}
\acrodef{ReLU}{rectified linear unit}

\begin{document}

\title{Tree Meets Transformer: A  Hybrid Architecture for Scalable Power Allocation in Cell-Free Networks}
 \author{\IEEEauthorblockN{Irched Chafaa\,\orcidlink{0000-0003-1467-5933},  Giacomo Bacci\,\orcidlink{0000-0003-1762-8024}, and Luca Sanguinetti\,\orcidlink{0000-0002-2577-4091}\thanks{This work was supported by the Smart Networks and Services Joint Undertaking (SNS JU) under the European Union’s Horizon Europe research and innovation program under Grant Agreement No 101192369 (6G-MIRAI), by the Italian Ministry of Education and Research (MUR) in the framework of the FoReLab Project (Departments of Excellence), by the HORIZON-JU-SNS-2022 EU project TIMES under grant no. 101096307, and by the European Union under the Italian National Recovery and Resilience Plan (NRRP) of NextGenerationEU, partnership on ``Telecommunications of the Future'' (PE00000001 -- Program ``RESTART'', Structural Project 6GWINET, Cascade Call SPARKS).}} 
\IEEEauthorblockA{\small \textit{  
Dipartimento di Ingegneria dell'Informazione, University of Pisa, 56122 Pisa, Italy} \\
irched.chafaa@ing.unipi.it, \{giacomo.bacci, luca.sanguinetti\}@unipi.it  }
}

\maketitle

\begin{abstract}
Power allocation remains a fundamental challenge in wireless communication networks, particularly under dynamic user loads and large-scale deployments. While Transformer-based models have demonstrated strong performance, their computational cost scales poorly with the number of users. In this work, we propose a novel hybrid Tree-Transformer architecture that achieves scalable  per-user power allocation. Our model compresses user features via a binary tree into a global root representation, applies a Transformer encoder solely to this root, and decodes per-user uplink and downlink powers through a shared  decoder. This design achieves logarithmic depth and linear total complexity, enabling efficient inference across large and variable user sets without retraining or architectural changes. We evaluate our model on the max-min fairness problem in cell-free massive MIMO systems and demonstrate that it achieves near-optimal performance while significantly reducing inference time compared to full-attention baselines. 
\end{abstract} 


\acresetall

\section{Introduction}

Power allocation is a critical task in wireless  networks, particularly in cell-free \ac{mMIMO} systems where users are served by distributed \acp{AP} without cell boundaries \cite{ngo2017cell}. The objective is to allocate \ac{UL} and \ac{DL} transmission power efficiently to maximize system throughput and fairness among \acp{UE}, while adapting to dynamic user loads and spatial configurations.

\subsection{Related Work}
Traditional optimization algorithms \cite{farooq2020accelerated, demir2021foundations, miretti2022closed}, such as convex solvers and iterative heuristics, provide reliable solutions but suffer from high computational complexity and poor scalability in large and dynamic networks. Deep learning models \cite{kim2023survey,mao2018deep} have emerged as promising alternatives, enabling fast inference and generalization across varying scenarios. However, neural architectures for power allocation rely on fixed input dimensions, making them inflexible to changes in the number of \acp{UE} and \acp{AP}. This dependency requires retraining or architectural redesign whenever the wireless network size varies, limiting their applicability in real-world deployments.

Transformer-based architectures \cite{kocharlakota2024pilot, chafaa2025transformer} have recently shown strong performance due to their ability to capture global dependencies across \acp{UE} and \acp{AP}, while offering flexibility with respect to network size during both training and inference. However, their quadratic complexity with respect to sequence length (number of \acp{UE}/\acp{AP}) \cite{vaswani2017attention} remains a bottleneck for large-scale deployments. To address this, efficient Transformer variants with linear attention complexity, such as CosFormer and Performer, have been proposed \cite{elouargui2023comprehensive}.

In our latest work \cite{chafaa2025linear}, we introduced a modified CosFormer architecture to jointly predict transmission powers and \ac{AP} clusters serving each \ac{UE}. While this model achieves linear complexity with respect to the number of \acp{UE}, it still requires full attention computation across all users. Moreover, its reliance on dense inter-user attention makes it less interpretable and harder to deploy in modular or distributed settings.

\subsection{Contributions}
To overcome these limitations, we propose a novel hybrid Tree-Transformer architecture for scalable and expressive per-user power allocation. Our model compresses user features into a global root representation using a binary tree structure \cite{harer2019tree}, applies a Transformer encoder \cite{vaswani2017attention} solely to this root, and decodes \ac{UL} and \ac{DL} power levels for each user via a shared  decoder. Through this design, the model fuses local user features with the global state of the wireless network, enabling accurate per‑user power predictions that generalize well across varying network sizes and layouts.

We evaluate our model on the max-min fairness problem in cell-free \ac{mMIMO} systems and compare it against full-attention baselines, including the standard Transformer, CosFormer and Performer-based models. Our results show that the hybrid model achieves near-optimal performance while significantly reducing inference time and computation complexity.

In summary, the main contributions of this paper are as follows:
\begin{itemize}
    \item We introduce a hybrid Tree-Transformer architecture that combines binary tree compression, root-level attention, and shared decoding for scalable power allocation.
    \item We demonstrate that the model achieves logarithmic depth and linear total complexity, with Transformer cost independent of user count.
    \item We show that the model generalizes to varying network settings, and supports modular, interpretable deployment.
    \item We validate the model on various scenarios, achieving near-optimal performance with significantly reduced computational overhead.
\end{itemize}

 \section{System Model and Problem Formulation}\label{sec:systemModel}
\begin{figure}[t]
  \centering
 \includegraphics[width=0.9\columnwidth]{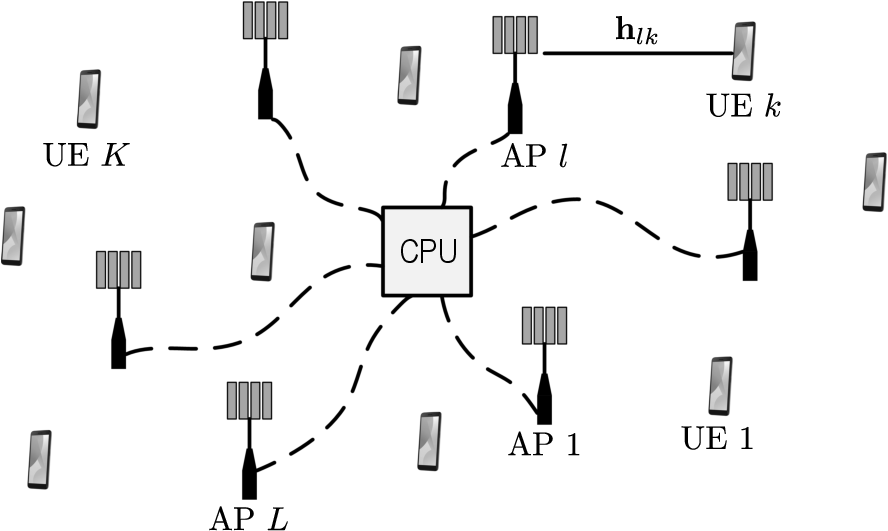}
  \caption{Illustration of a  cell-free \acs{mMIMO} network. Users are served by \acp{AP} without cell boundaries.}
  \label{fig:cfmMIMO}
\end{figure}

We consider a cell-free \ac{mMIMO} system (\figurename~\ref{fig:cfmMIMO}), where $K$ single-antenna \acp{UE} are served by $L$ \acp{AP} with $N$ antennas each. The \acp{AP} coordinate via a fronthaul network and a \ac{CPU} for joint processing and power allocation. The standard \ac{TDD} protocol of cell-free \ac{mMIMO} is used \cite{demir2021foundations}, where the $\tau_c$
available channel uses are employed for: 
\begin{enumerate*}[label=(\emph{\roman*})]
\item \ac{UL} training phase ($\tau_p$); 
\item \ac{DL} payload transmission ($\tau_d$); and 
\item \ac{UL} payload transmission ($\tau_u$).
\end{enumerate*}
 Clearly, $\tau_c\geq\tau_p + \tau_d + \tau_u$.

 We consider a narrowband channel model and assume that the channel remains constant within a coherence block. We denote the channel vector between the \ac{AP} $l$ and \ac{UE} $k$ with $\mathbf{h}_{lk}$, and model it as \cite{demir2021foundations}:
\begin{align}\label{eq:channel_model}
  \mathbf{h}_{lk} = \sqrt{\beta_{lk}} \mathbf{R}_{lk}^{1/2} \mathbf{g}_{lk},
\end{align}
where $\beta_{lk}$ is the large-scale fading, accounting for path loss and shadowing, $\mathbf{R}_{lk} \in \mathbb{C}^{N \times N}$ is the spatial correlation matrix at \ac{AP} $l$, and $\mathbf{g}_{lk} \sim \mathcal{N}_C(\mathbf{0}, \mathbf{I}_N)$ is an i.i.d. complex Gaussian vector representing the small-scale fading, where $\mathbf{I}_N$ is the $N \times N$ identity matrix. We assume that the channels $\{\mathbf{h}_{lk}; l=1,\ldots,L\}$ are independent and call $\mathbf{h}_{k} = \left[\mathbf{h}_{1k}^\Ttran, \ldots, \mathbf{h}_{Lk}^\Ttran \right]^\Ttran \in \mathbb{C}^{LN}$ the collective channel from all \acp{AP} to \ac{UE} $k$.

The \ac{CPU} computes the estimate of $\mathbf{h}_{k}$ on the basis of received pilot sequences transmitted during the training phase \cite{demir2021foundations}. The \ac{MMSE} estimate is $\widehat{\mathbf{h}}_{k} = [\widehat{\mathbf{h}}_{1k}^\Ttran, \ldots, \widehat{\mathbf{h}}_{Lk}^\Ttran ]^\Ttran$ with \cite{demir2021foundations}
\begin{align}
\widehat{\mathbf{h}}_{lk} = \mathbf{R}_{lk} \mathbf{Q}_{lk}^{-1} \left( \mathbf{h}_{l k} +  \frac{1}{\tau_p \rho} \mathbf{n}_{lk} \right) \sim \mathcal{N}_C \left( \mathbf{0}_N, \mathbf{\Phi}_{lk} \right),
\end{align}
where $\rho$ is the \ac{UL} pilot power of each \ac{UE}, $\mathbf{n}_{lk} \sim \mathcal{N}_C(\mathbf{0}_{N}, \sigma^2\mathbf{I}_{N})$ is the thermal noise, and $\mathbf{\Phi}_{lk} = \mathbf{R}_{lk} \mathbf{Q}_{lk}^{-1} \mathbf{R}_{lk}$, where $\mathbf{Q}_{lk} = \mathbf{R}_{l k} + \frac{\sigma^2}{\tau_p\rho} \mathbf{I}_{N}$. Hence, $\widehat{\mathbf{h}}_{k} \sim \mathcal{N}_C \left( \mathbf{0}_{LN}, \mathbf{\Phi}_{k} \right)$, with $\mathbf{\Phi}_{k} = {\mathrm{diag}}(\mathbf{\Phi}_{1k},\ldots,\mathbf{\Phi}_{Lk})$. Note that the method proposed in this paper can be applied to other channel estimation schemes. 

\subsection{Uplink and Downlink Transmissions}
To detect the data of \ac{UE} $k$ in the \ac{UL}, the \ac{CPU} selects an arbitrary receive combining vector $\mathbf{v}_{k}\in \mathbb{C}^{LN}$ based on all the collective channel estimates $\{\widehat{\mathbf{h}}_{k}; k =1,\ldots, K\}$. An achievable \ac{SE} of \ac{UE} $k$ is given by \cite{demir2021foundations}: 
\begin{align}\label{eq:spectral_efficiency_uplink}
  \text{\ac{SE}}_k^\text{UL} = \frac{\tau_u}{\tau_c} \log_2(1 + \text{SINR}_k^\text{UL}),
\end{align}
with the effective \ac{SINR} defined as
 \begin{align}
\!\!\!\!\!\frac{p_k^\text{UL} \left| \mathbb{E} \left\{ \mathbf{v}_{k}^\Htran \mathbf{h}_{k} \right\} \right|^2}{\sum\limits_{i=1}^K p_i^\text{UL} \mathbb{E} \left\{ \left| \mathbf{v}_{k}^\Htran \mathbf{h}_{i} \right|^2 \right\} - p_k^\text{UL} \left| \mathbb{E} \left\{ \mathbf{v}_{k}^\Htran \mathbf{h}_{k} \right\} \right|^2 + \sigma^2\mathbb{E} \left\{ \| \mathbf{v}_k \|^2 \right\}},
\end{align}
where $p_k^\text{UL}$ is \ac{UE} $k$'s \ac{UL} transmit power. The expectation $\mathbb{E}\{\cdot\}$ is taken with respect to all sources of randomness. Although the bound in \eqref{eq:spectral_efficiency_uplink} is valid for any combining vector, we consider the \ac{MMSE} combiner, given by\cite{demir2021foundations}: 
\begin{align}
\mathbf{v}_{k} = \left( \sum_{k=1}^{K} p_k^\text{UL} \widehat{\mathbf{h}}_{k} \widehat{\mathbf{h}}_{k}^\Htran + \mathbf{Z} \right)^{-1} \widehat{\mathbf{h}}_{k}, 
\end{align}
where $\mathbf{Z} =  \sum_{k=1}^{K} p_k^\text{UL}\left[\mathrm{diag}(\mathbf{R}_{1k},\ldots,\mathbf{R}_{Lk}) - \mathbf{\Phi}_{k}\right] + \sigma^2 \mathbf{I}_{LN}$.

In the \ac{DL}, the \ac{CPU} coordinates the \acp{AP} to transmit signals to the \acp{UE}. Similarly to \ac{UL}, an achievable \ac{SE} of user $k$ is obtained as:
\begin{align}
  \text{ \ac{SE}}_k^\text{DL} = \frac{\tau_d}{\tau_c}\log_2(1 + \text{SINR}_k^\text{DL}),
\end{align}
with the effective \ac{SINR} defined as
 \begin{align}
\frac{p_k^\text{DL} \left| \mathbb{E} \left\{ \mathbf{h}_{k}^\Htran \mathbf{w}_{k} \right\} \right|^2}{\sum\limits_{i=1}^K p_i^\text{DL} \mathbb{E} \left\{ \left| \mathbf{h}_{k}^\Htran \mathbf{w}_{i} \right|^2 \right\} - p_k^\text{DL} \left| \mathbb{E} \left\{ \mathbf{h}_{k}^\Htran \mathbf{w}_{k} \right\} \right|^2 + \sigma^2},
\end{align}
where $p_{k}^\text{DL}$ is the \ac{DL} power used by the \ac{CPU} to serve \ac{UE} $k$ such that $p_{k}^\text{DL}=\sum_{l=1}^{L}{p_{k,l}^\text{DL}}$, with $p_{k,l}^\text{DL} \in \mathclose{[}0,\overline{P}_l^\text{DL}\mathclose{]}$ being the \ac{AP} $l$'s transmit power allocated for user $k$; $\overline{P}_l^\text{DL}$ is the maximum power per \ac{AP}; and $\mathbf{w}_{k}\in \mathbb{C}^{LN}$ is the associated unit-norm precoding vector. The \ac{MMSE} precoder is used \cite{demir2021foundations}, which is given by $\mathbf{w}_{k} = \mathbf{v}_{k} / \|\mathbf{v}_{k}\|$.

\subsection{Problem Formulation}
Our goal is to allocate transmission powers $\{p_k^\text{UL}, p_k^\text{DL}\}$ to ensure SE fairness among users while satisfying system constraints. Specifically, we consider the max-min fairness criterion, which aims to maximize the minimum achievable \ac{SE} across all \acp{UE}. This leads to the following optimization problem \cite{demir2021foundations}:
\begin{align}
\begin{aligned}
  \max_{\{p_k^\text{UL}\geq 0\}} \min_k &\ \textrm{\ac{SE}}_k^\text{UL} \\
  \text{subject to} &\quad  p_k^\text{UL} \leq \overline{P}_k^\text{UL} \ \forall k
\end{aligned}
\label{optul}
\end{align}
where $\overline{P}_k^\text{UL}$ is the maximum \ac{UL} power for user $k$. Similarly, in the \ac{DL} we have that:
\begin{align}
\begin{aligned}
  \max_{\{p_k^\text{DL}\geq 0\}} \min_k &\ \text{\ac{SE}}_k^\text{DL}\\
  \text{subject to} &\quad \textstyle{\sum_{k=1}^K p_k^\text{DL} \leq \sum_{l=1}^L \overline{P}_l^\text{DL}}
\end{aligned}
\label{optdl}
\end{align}
where the constraint ensures that the total power allocated to all \acp{UE} does not exceed the total power budget across all \acp{AP}.

Both optimization problems can be solved using closed-form approximations \cite{miretti2022closed}, iterative solvers \cite{farooq2020accelerated, demir2021foundations}, or deep learning approaches \cite{kim2023survey, mao2018deep, choi2025deep, kocharlakota2024pilot}. However, as discussed earlier, these methods either suffer from high computational cost, require retraining for different network sizes, or scale poorly with the number of users. To address these limitations, we propose a hybrid Tree-Transformer architecture that learns to predict \ac{UL} and \ac{DL} power directly from \ac{UE} and \ac{AP} positions. 

\section{Proposed Hybrid Tree-Transformer Model}

 This section details the proposed learning approach for power prediction including: the dataset construction,  input processing, model design, and training procedure.

\subsection{Dataset Construction}

To train and evaluate the proposed model, we construct a synthetic dataset using typical simulation parameters for cell-free \ac{mMIMO} systems \cite{demir2021foundations}. 
   
\begin{enumerate}[label=\emph{\alph*})]
    \item \emph{Data generation}. For each configuration defined by a pair $(K, L)$, we generate:
    \begin{itemize}
        \item random \ac{2D} positions in a given area for $K$ single-antenna \acp{UE}, denoted as $\{\mathbf{u}_k\}_{k=1}^K$, with $\mathbf{u}_k\in\mathbb{R}^2$, and $L$ \acp{AP}, each equipped with $N$ antennas, denoted as $\{\mathbf{a}_l\}_{l=1}^L$, with $\mathbf{a}_l\in\mathbb{R}^2$;
        \item channel realizations $\mathbf{h}_{lk}$ based on the model in \eqref{eq:channel_model}, using spatial correlation matrices and large-scale fading coefficients $\beta_{lk}$;
        \item optimal \ac{UL} and \ac{DL} power allocations computed using the closed-form max-min fairness solution from \cite{miretti2022closed}.
    \end{itemize}

    \item \emph{Noise Injection}. To simulate realistic deployment conditions and improve generalization, we inject Gaussian noise into the \ac{UE}  positions:
    \begin{align}
        \label{eq:tilde_u}
        \tilde{\mathbf{u}}_k &= \mathbf{u}_k + \delta_k, \quad \delta_k \sim \mathcal{N}(0, \sigma_k^2),
    \end{align}
    where $\sigma_k  =5\meter$ controls the noise level\cite{xue2021analysis}.


    \item \emph{Normalization}. All features (positions and powers) are normalized to the range $[0, 1]$ using min-max scaling \cite{mao2018deep}. For a feature vector $\boldsymbol\xi$, the normalized value is computed as
    \begin{align}
        \boldsymbol{\xi}_{\text{norm}} = \frac{\boldsymbol{\xi} - \min(\boldsymbol{\xi})}{\max(\boldsymbol{\xi}) - \min(\boldsymbol{\xi}) + \varepsilon},
    \end{align}
    where $\varepsilon$ is a small constant added to avoid division by zero. Each \ac{2D} position with spatial coordinates $(x, y)$ is normalized by applying min-max scaling separately to the $x$- and $y$-coordinates across all \acp{AP} or \acp{UE}. This preserves geometric relationships such as distances and angles between nodes while standardizing the input range for stable training. \ac{UL} and \ac{DL} powers are normalized globally across all samples to ensure consistent scaling.

    \item \emph{Dataset overview}. A total number of $8000$  samples  for each value of $K \in \{2,4,6,8,10\}$ and $L =16$ are used for training. This diversity enables the model to learn robust mapping across varying network sizes, spatial layouts and channel realizations. Each sample includes \ac{UE} and \ac{AP} positions as input, and optimal power allocations as output labels. Additionally, $200$ samples for each value of $K= 2, 3, \ldots, 40$ and $L\in  \{1,4,9,16,25\}$ are used for testing to assess the model's ability to generalize to new configurations.
        
\end{enumerate}


\subsection{Model Architecture}

\begin{figure}[t]
  \centering
 \includegraphics[width=\columnwidth]{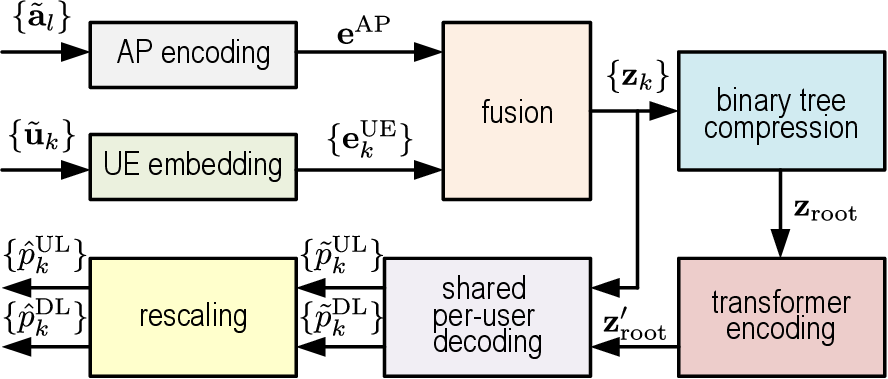}
  \caption{Block diagram of the proposed model architecture.}
  \label{fig:model_architecture}
  \end{figure}

The proposed hybrid Tree-Transformer model is designed to predict \ac{UL} and \ac{DL} power allocations for each UE based on the spatial configuration of UEs and \acp{AP}. The architecture consists of seven main components: an \ac{AP} encoder, a \ac{UE} embedding module, a fusion method, a binary tree compressor, a Transformer encoder applied to the root node, a shared per-user decoder, and a final rescaler. The overall structure is illustrated in \figurename~\ref{fig:model_architecture}.

\begin{enumerate}[label=\emph{\alph*})]
    \item \emph{\ac{AP} encoder}. The coordinates of all \acp{AP} $\{\tilde{\mathbf{a}}_l\}_{l=1}^L$ are passed through a two-layer \ac{MLP}. The first linear layer maps the \ac{2D} inputs to a hidden dimension $d_\text{enc}$, followed by a \ac{ReLU} activation \cite{rasamoelina2020review}. A second linear layer refines the representation while keeping the same size. Hence, the embedding of the $l$-th \ac{AP} becomes $\mathbf{e}_l^\text{AP} = \text{MLP}(\tilde{\mathbf{a}}_l) \in \mathbb{R}^{d_\text{enc}}$, whereas the global \ac{AP} context is obtained by averaging across them, as $\mathbf{e}^\text{AP} = \frac{1}{L} \sum_{l=1}^L \mathbf{e}_l^\text{AP}$. This global vector $\mathbf{e}^\text{AP}\in \mathbb{R}^{d_\text{enc}}$ summarizes the whole distribution of \acp{AP} in the environment.

    \item \emph{\ac{UE} embedding}. Similarly to the \acp{AP}, each \ac{UE} is embedded via the linear transformation $\mathbf{e}_k^\text{UE} = \mathbf{W}_\text{UE} \tilde{\mathbf{u}}_k + \mathbf{b}_\text{UE}$, where $\mathbf{W}_\text{UE}\in\mathbb{R}^{d_\text{enc}\times2}$ and $\mathbf{b}_\text{UE}\in\mathbb{R}^{d_\text{enc}}$ are trainable parameters.

    \item \emph{Fusion}. The \ac{AP} embedding $\mathbf{e}^\text{AP}$ is concatenated to each user embedding $\mathbf{e}_k^\text{UE}$ to incorporate global context: $\mathbf{z}_k = [(\mathbf{e}_k^\text{UE})^\Ttran\, (\mathbf{e}^\text{AP})^\Ttran]^\Ttran \in \mathbb{R}^{2d_\text{enc}}$. The set of fused descriptors $\{\mathbf{z}_k\}_{k=1}^K$  serves as the input to both the binary tree compressor and the shared decoder.

    \begin{figure}[t]
      \centering
     \includegraphics[width=0.9\columnwidth]{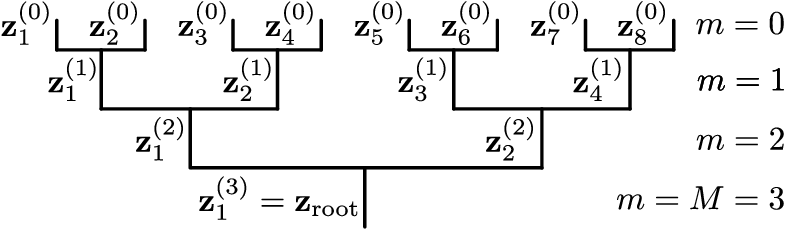}
      \caption{Example of binary tree compression ($K=8$).}
      \label{fig:treeExample}
      \vspace*{-0.3cm}

    \end{figure}

\item \emph{Binary tree compressor}. 
The fused user descriptors $\{\mathbf{z}_k\}_{k=1}^K$ are first projected into a common embedding space of dimension $d_\text{mod}$ using a linear transformation $\mathbf{z}_k^{(0)} = \mathbf{W}_0 \mathbf{z}_k + \mathbf{b}_0$, where $\mathbf{W}_0 \in \mathbb{R}^{d_\text{mod}\times 2d_\text{enc}}$ and $\mathbf{b}_0 \in \mathbb{R}^{d_\text{mod}}$. 
They are then hierarchically aggregated using a binary tree structure with $M=\lceil\log_2K\rceil$ compression stages, where $\lceil\cdot\rceil$ is the ceiling function, as illustrated in \figurename~\ref{fig:treeExample} for $K=8$ ($M=3$). 
At each stage $m=1,\dots,M$, the parent vectors $\mathbf{z}_q^{(m)}$, $q=1,\dots, 2^{M-m}$, are obtained using the linear transformation $\mathbf{z}_q^{(m)}=\mathbf{W}[(\mathbf{z}_{2q-1}^{(m)})^\Ttran\,(\mathbf{z}_{2q}^{(m)})^\Ttran]^\Ttran+\mathbf{b}$, where $\mathbf{b}\in\mathbb{R}^{d_\text{mod}}$ is a trainable bias, and $\mathbf{W}$ is a learnable weight matrix, with size $d_\text{mod}\times 2 d_\text{mod}$.
If the number of descriptors at any level is odd, a zero-padding vector is appended to enable pairwise merging. 
The process continues until a single root descriptor $\mathbf{z}_\text{root} = \mathbf{z}_1^{(M)}$ is obtained. 
Since the depth of the tree is logarithmic in $K$, and the total number of merge operations scales linearly with $K$, the compression is efficient and scalable. 
This hierarchical aggregation strategy is inspired by prior work in tree-based encoders \cite{tai2015improved,harer2019tree}, enabling global context extraction without quadratic complexity.

\item \emph{Transformer encoder}. The root descriptor $\mathbf{z}_\text{root}$ is further refined using a Transformer encoder \cite{vaswani2017attention} with $S$ stacked layers, each comprising a multi-head self-attention block followed by a feed-forward network. Each attention block uses $A$ attention heads, and all internal representations maintain the same dimensionality $d_\text{mod}$. 

Since the encoder operates on a single root descriptor rather than a sequence of tokens, the self-attention mechanism degenerates to multi-head linear projections of this vector. In practice, each head applies a distinct learned transformation, and their concatenation enriches the global representation. This design allows the model to benefit from the expressive diversity of multi-head attention while avoiding the quadratic cost of conventional Transformers. The resulting enriched global context vector $\mathbf{z}_\text{root}^\prime \in \mathbb{R}^{d_\text{mod}}$ is then used by the decoder at the next stage.

    \item \emph{Shared per-user decoder}. The enriched global context vector $\mathbf{z}_\text{root}^\prime$ is broadcast to all \acp{UE} and concatenated with their original fused descriptors $\{\mathbf{z}_k\}_{k=1}^{K}$. This results in a context-aware vector for each user, $\mathbf{z}_k^\text{dec} = [(\mathbf{z}_k)^\Ttran\, (\mathbf{z}_\text{root}^\prime)^\Ttran]^\Ttran \in \mathbb{R}^{2d_\text{enc} + d_\text{mod}}$. Each $\mathbf{z}_k^\text{dec}$ is then passed through a shared decoder network, implemented as a two-layer \ac{MLP} with \ac{ReLU} activation, followed by a sigmoid output \cite{rasamoelina2020review}. The decoder maps the vector $\mathbf{z}_k^\text{dec}$ to a pair of normalized power values $\tilde{\mathbf{p}}_k =  [\tilde{p}_k^\text{UL}, \tilde{p}_k^\text{DL}] \in [0, 1]^2$, where $\tilde{p}_k^\text{UL}$ and $\tilde{p}_k^\text{DL}$ denote the normalized predicted \ac{UE} $k$'s \ac{UL} and \ac{DL} powers. This shared decoding strategy enables parameter efficiency and consistent prediction behavior across users, regardless of the network size $(K,L)$. By conditioning each prediction on both local user features and the globally refined context, the model captures both fine-grained and holistic patterns in the wireless network for power prediction.

    \item \emph{Rescaler}. A final step involves rescaling the predictions back to their original ranges, by applying  de-normalization and rescaling:
    \begin{align}
        \label{eq:hat_UL}
        \hat{p}_k^\text{UL} &= \Delta_\text{UL}\tilde{p}_k^\text{UL} + \underline{P}_\text{UL}, \\
        \label{eq:hat_DL}
        \hat{p}_k^\text{DL} &= \frac{\check{p}_k^\text{DL}\sum_{l=1}^L \overline{P}_l^\text{DL}}{\sum_{k=1}^K \check{p}_k^\text{DL}}, \textrm{with
        $\check{p}_k^\text{DL}\!=\!\Delta_\text{DL}\tilde{p}_k^\text{DL}\!+\!\underline{P}_\text{DL}$},
    \end{align}
    where $\Delta_\text{UL}=\overline{P}_\text{UL}-\underline{P}_\text{UL}$ (resp., $\Delta_\text{DL}=\overline{P}_\text{DL}-\underline{P}_\text{DL}$) is the \ac{UL} (resp., \ac{DL}) power range, with $\overline{P}_\text{UL}$ (resp. $\underline{P}_\text{UL}$) denoting \ac{UE} $k$'s maximum (resp., minimum) \ac{UL} power, and $\overline{P}_\text{DL}$ (resp. $\underline{P}_\text{DL}$) being the counterpart on the \ac{DL}.    

\end{enumerate}

\subsection{Model Parameters}

The hybrid Tree-Transformer model is implemented using PyTorch \cite{paszke2019pytorch}, and configured with the parameters in \tablename~\ref{tab:model_params}, selected empirically to balance model expressiveness and computational efficiency. The architecture remains lightweight, enabling scalability across varying user counts while retaining sufficient capacity to learn complex mappings from spatial features to power allocation.

\begin{table}[t]
\centering
\resizebox{\columnwidth}{!}{%
\begin{tabular}{|l|c|l|c|}
\hline
\multicolumn{2}{|c|}{\textbf{model parameters}} & \multicolumn{2}{c|}{\textbf{training parameters}} \\ \hline
\acs{AP}/\acs{UE} embedding dimension $d_\text{enc}$ & $32$ & learning rate & $10^{-3}$ \\ \hline
model dimension $d_\text{mod}$ & $64$ & batch size  & $128$ \\ \hline
number of attention heads $A$ & $4$ & number of epochs & $100$ \\ \hline
number of Transformer layers $S$ & $2$ & optimizer & AdamW \cite{zhou2024towards} \\ \hline
\end{tabular}
}
\caption{Model parameters and training setup configuration.}
\label{tab:model_params}
\end{table}

\subsection{Training Setup}
The hybrid Tree-Transformer model is trained to minimize the \ac{MSE} $\mathcal{L}_\text{MSE}$ between the normalized predicted powers  and the optimal ones obtained from the closed-form solution \cite{miretti2022closed}.
By doing so, the model implicitly learns the effects of the  channel propagation environment, since the optimal powers embed channel information. As a result, the network is trained to approximate the mapping from the spatial positions of \acp{UE} and \acp{AP} to power allocations that maximize the minimum \ac{SE}.

The training is performed across multiple values of the number of users $K$, ensuring that the model is exposed to heterogeneous scenarios and can generalize across different network sizes. Validation is also conducted across varying sizes $(K,L)$. The specific hyperparameters used during training are summarized in \tablename~\ref{tab:model_params}. The best-performing model is saved based on the lowest average validation loss across different network configurations.


\begin{figure}[t]
  \centering
  \includegraphics[width=\columnwidth]{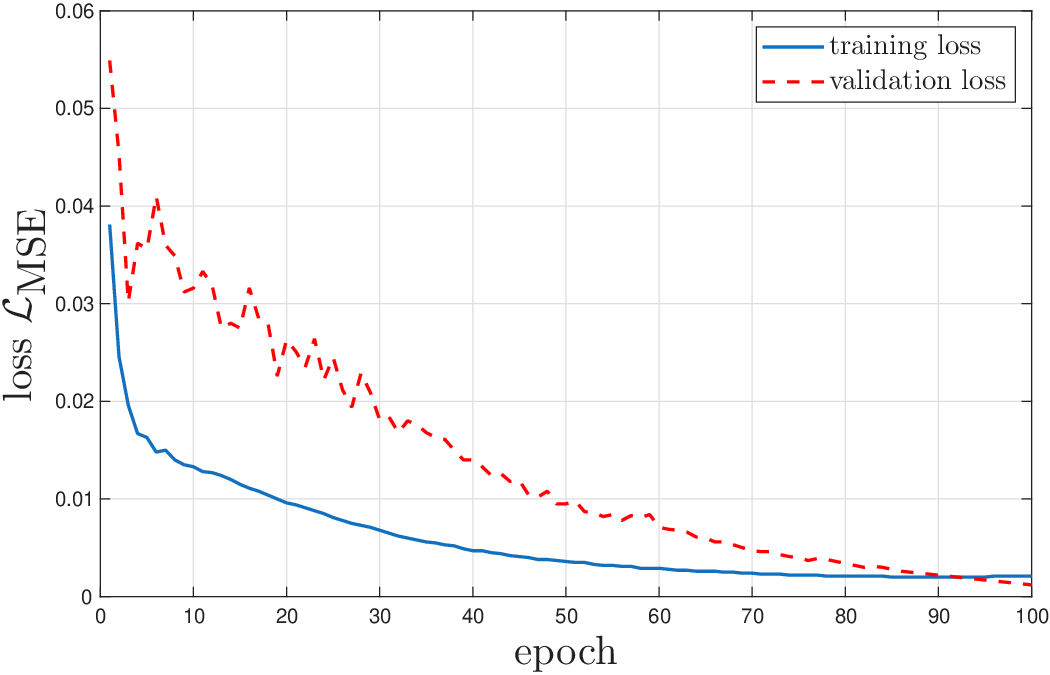}
  \caption{Evolution of training and validation loss. The model converges rapidly and generalizes well across network configurations.}
  \label{fig:loss_curves}
  \vspace*{-0.3cm}
\end{figure}

\section{Numerical Results}
In this section, we present numerical results to illustrate the performance of the proposed approach for a cell-free \ac{mMIMO} system, as described in \sectionname~\ref{sec:systemModel}. We consider a network with a coverage area of $500\meter \times 500\meter$, with $N = 4$ antennas per \ac{AP}. The \acp{AP} are uniformly deployed within the squared coverage area. The maximum \ac{UL} transmit power for each user is $\overline{P}_\text{UL}=100\mW$, whereas the maximum \ac{DL} transmit power for each \ac{AP} is $\overline{P}_l^\text{DL}=200\mW$. We assume $\tau_c = 200$ and set $\tau_p = 10$, $\tau_u = 90 $  and $\tau_d = 100$. Large-scale fading coefficients are computed following the 3GPP path-loss model adopted in \cite[\sectionname~III-D]{miretti2022closed} for a $2$-GHz carrier frequency, a pathloss exponent of $3.67$, a \ac{UE}-\ac{AP} height difference of $10\meter$ and a shadow fading $F_{kl} \sim \mathcal{N}_C(0, \alpha^2)$, with $ \alpha^2 = 4\dB$. The shadow fading terms are spatially correlated as in \cite[\sectionname~III-D]{miretti2022closed} to account for the fact that closely located \acp{UE} experience similar shadow fading effects. The noise power is $\sigma^2 = -94\dBm$ \cite{miretti2022closed} with a noise figure $\eta = 7\dB$ and a bandwidth $B = 20\MHz$. The dataset and codes for model training are available in \cite{hybridmodel}.

\emph{Training performance}. \figurename~\ref{fig:loss_curves} shows the evolution of training and validation loss $\mathcal{L}_\text{MSE}$ over epochs. The consistent downward trend in both curves indicates that the model effectively learns to generalize across varying user distributions. The eventual close alignment between training and validation loss suggests minimal overfitting and strong predictive performance. Minor fluctuations in the validation curve are attributed to variability in spatial layouts and channel conditions across samples. Nevertheless, the overall convergence behavior confirms the robustness of the learning process and the model’s ability to accurately predict power.

\begin{figure}[t]
  \centering
  \includegraphics[width=\columnwidth]{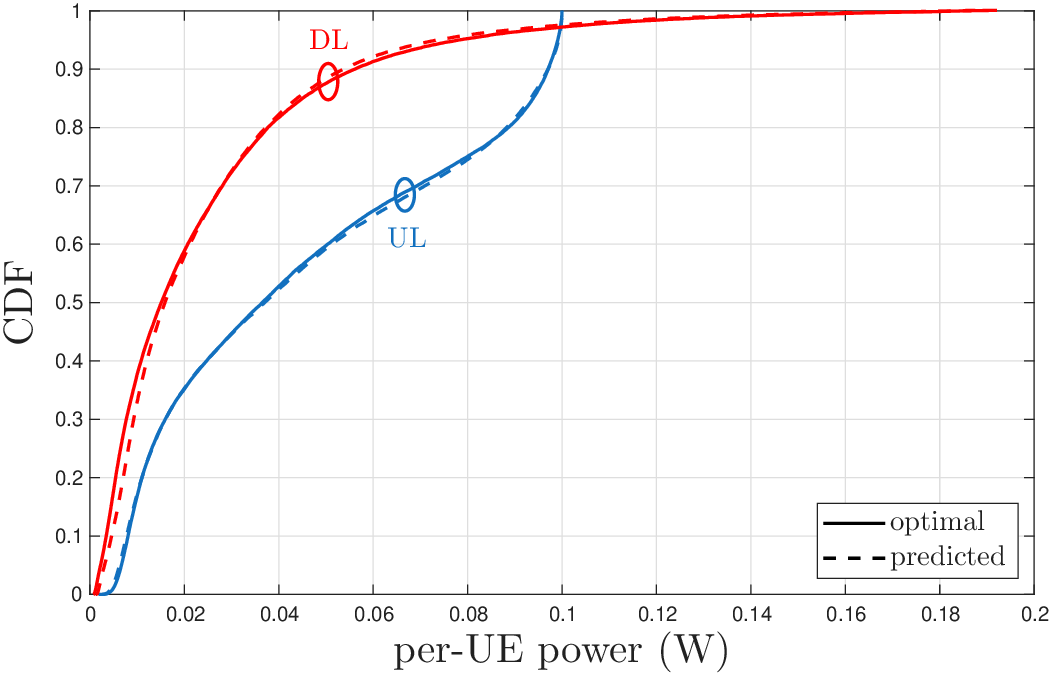}
  \caption{\acsp{CDF} of optimal and predicted power on the test set. The predicted curves closely follow the optimal ones, confirming the model’s robustness and generalization capability.}
  \label{fig:cdf_curves}
  \vspace*{-0.3cm}
\end{figure}

\emph{Prediction accuracy}. \figurename~\ref{fig:cdf_curves} shows the \acp{CDF} of the optimal and predicted \ac{UL} and \ac{DL} power on the test set, which includes diverse network configurations $(K,L)$ not encountered during training. We can see that the curves of predicted and optimal powers almost completely overlap, demonstrating that the  model can generalize effectively to new settings and reproduce the optimal \ac{UL} and \ac{DL} power distributions with high fidelity, even under noisy input conditions. 

\begin{figure}[htbp]
    \centering
    \begin{subfigure}[t]{\columnwidth}
        \centering
        \includegraphics[width=\columnwidth]{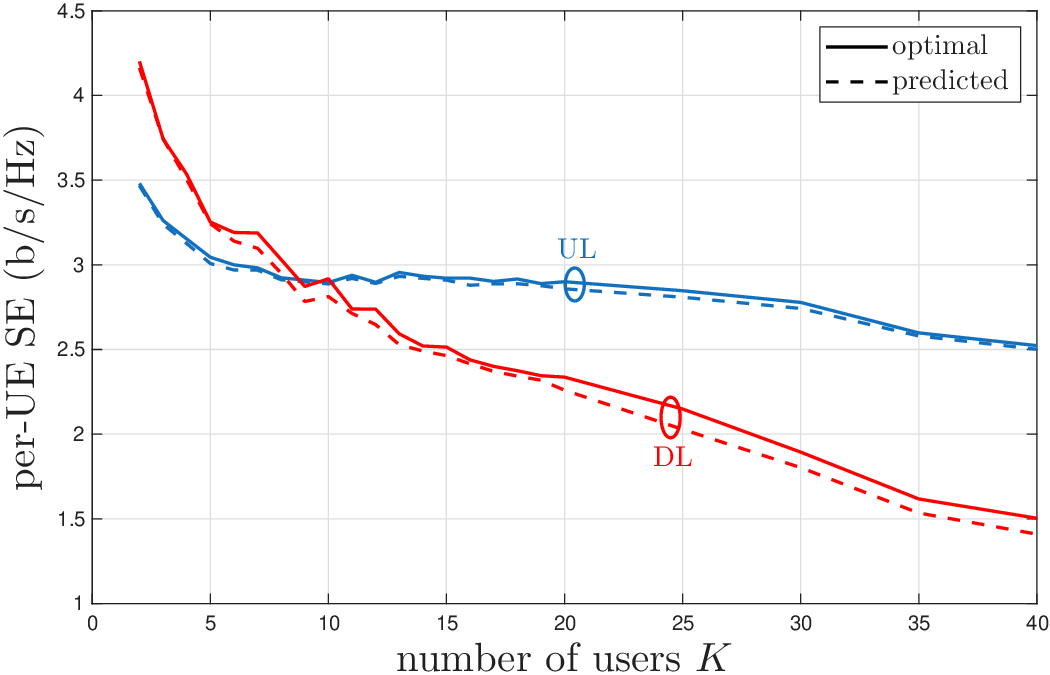} 
        \caption{Per-\acs{UE} \acs{SE} vs. number of \acsp{UE} $K$ with fixed $L=16$.}
        \label{fig:se_vs_K}
    \end{subfigure}
    \hfill
    \begin{subfigure}[t]{\columnwidth}
        \centering
        \includegraphics[width=\columnwidth]{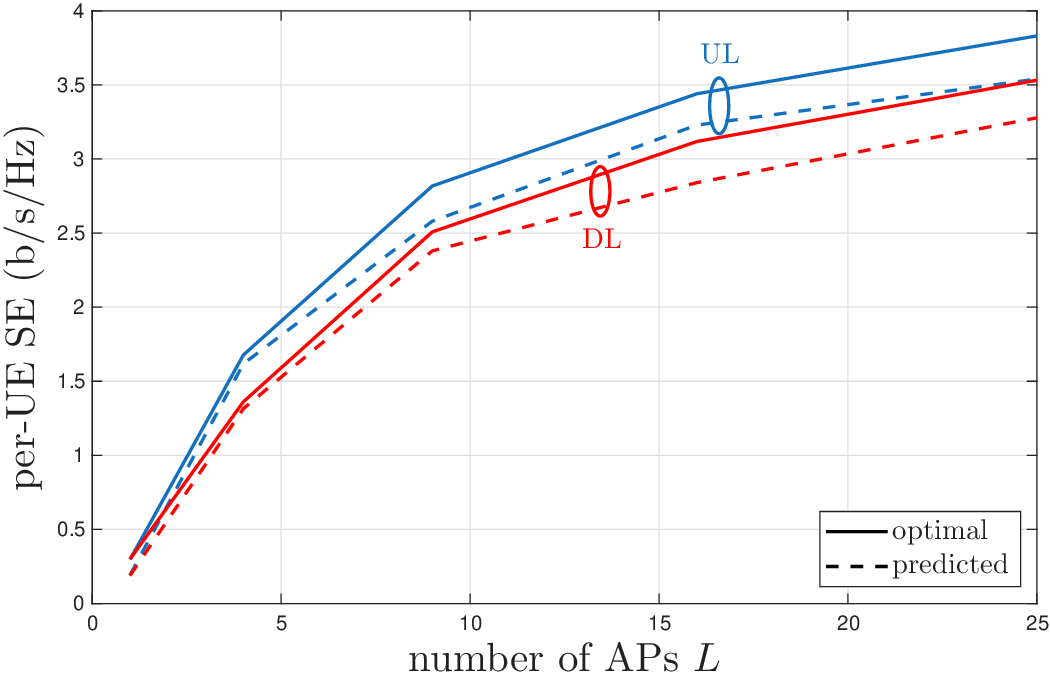} 
        \caption{Per-\acs{UE} \acs{SE} vs. number of \acsp{AP} $L$ with fixed $K=10$.}
        \label{fig:se_vs_L}
    \end{subfigure}
    
    \caption{Comparison of \acs{SE} trends: (\subref{fig:se_vs_K})~varying \acs{UE} count $K$, (\subref{fig:se_vs_L})~varying \acs{AP} count $L$. Predicted \acs{SE} closely tracks the optimal \acs{SE} across both scenarios.}
    \label{fig:se_vs_KL}
    \vspace*{-0.3cm}
\end{figure}

\emph{Flexibility with network size}. To assess the generalization capability of the proposed model, in \figurename~\ref{fig:se_vs_KL} we evaluate its \ac{SE} performance on new network settings with varying numbers of \acp{UE} $K$ and \acp{AP} $L$, without retraining or modifying the architecture. We can see that the predicted \ac{SE} curves closely follow the optimal ones with only a small gap: on average, the prediction remains within $0.12\bpsHz$ of the optimal \ac{SE} in both simulations, confirming that the model can adapt to different network sizes and configurations while maintaining high accuracy. Since the optimal closed-form solution is available and our model achieves comparable \ac{SE} values, we do not extend the \ac{SE} comparison to other methods in the literature.

\emph{Comparison with Transformer variants}. To highlight the efficiency of our proposed model, we compare it against  Transformer \cite{vaswani2017attention}, CosFormer, and Performer  \cite{elouargui2023comprehensive}. The results are reported in \tablename~\ref{tab:transformer_comparison} for $K=40$, $L=16$ (for the Performer case, we consider $F=256$ random features). For all variants, the input consists of \ac{UE}/\ac{AP} positions, and the output is the predicted \ac{UL} and \ac{DL} power, following the same setup as in \cite{chafaa2025transformer}.

Our model restricts attention computation to the compressed binary tree representation, yielding logarithmic depth and linear complexity. Unlike other models, its cost is independent of the wireless network size $(K,L)$. As a result, it achieves competitive \ac{SE} while offering the lowest latency among all compared models, as highlighted in bold in \tablename~\ref{tab:transformer_comparison}. This confirms its suitability for large-scale cell-free \ac{mMIMO} systems and distributed deployments where local prediction is required. As a reference, the optimal closed-form solution \cite{miretti2022closed} has complexity $\mathcal{O}(K^3)$, which is significantly higher than all learning-based approaches.

\begin{table*}[t]
\centering
\begin{tabular}{|l|c|c|c|c|}
\hline
\emph{metric} & \emph{our model} & \emph{Transformer} & \emph{CosFormer} & \emph{Performer} \\ \hline
\emph{parameters} & $584,354$ & $845,826$ & $102,338$ & $102,338$ \\ \hline
\emph{latency (ms)} & $\mathbf{3.57}$ & $28.46$ & $7.11$ & $25.45$ \\ \hline
\emph{avg. \acs{SE} (b/s/Hz)} & $\mathbf{2.01}$ & $1.66$ & $1.80$ & $1.81$ \\ \hline
\emph{scalability with $K$} & $\boldsymbol{\mathcal{O}\!\left(K \cdot d_\text{mod}^{2}\right)}$
 & $\mathcal{O}\!\left(K^2 \cdot S \cdot d_\text{mod}^{2}\right)$ & $\mathcal{O}\!\left(K \cdot S \cdot d_\text{mod}^{2}\right)$ & $\mathcal{O}\!\left(K \cdot S \cdot  d_\text{mod} \cdot F\right)$
 \\ \hline
\emph{scalability with $L$} & $\boldsymbol{\mathcal{O}\!\left(L \cdot d_\text{enc}\right)}$
 & $\mathcal{O}\!\left( L \cdot K \cdot d_\text{mod}\right)$ & $\mathcal{O}\!\left( L \cdot K \cdot d_\text{mod}\right)$ & $\mathcal{O}\!\left( L \cdot K \cdot d_\text{mod}\right)$
 \\ \hline
\end{tabular}
\caption{Performance comparison of power prediction models using  Transformer variants.}
\label{tab:transformer_comparison}
\vspace*{-0.3cm}
\end{table*}

\section{Conclusion}
In this work, we propose a Tree-Transformer architecture for power allocation in cell-free \ac{mMIMO} systems. By combining binary tree compression with Transformer-based global reasoning, the model achieves high prediction accuracy while significantly reducing complexity compared to both standard Transformer variants and the optimal closed-form solution. Numerical results demonstrate that the proposed approach generalizes well to unseen network configurations, remains robust under noisy inputs, and achieves SE values close to the optimal ones with minimal latency. Beyond the immediate performance gains, the Tree-Transformer offers scalability with respect to both the number of \acp{UE} and \acp{AP}, making it suitable for large-scale deployments and distributed implementations. 

As a perspective, a promising extension of this work is the integration of distributed learning strategies, where multiple \acp{AP} collaboratively train or update local models without centralized coordination. This approach would further enhance scalability and adaptability in large-scale deployments.

\bibliographystyle{IEEEtran}
\bibliography{biblio}

\end{document}